# Self-modulation instability of a long proton bunch in plasmas

Naveen Kumar* and Alexander Pukhov

*Institut für Theoretische Physik I, Heinrich-Heine-Universität Düsseldorf, D-40225 Germany*

Konstantin Lotov

*Budker Institute of Nuclear Physics and Novosibirsk State University, 630090 Novosibirsk, Russia*

An analytical model for the self-modulation instability of a long relativistic proton bunch propagating in uniform plasmas is developed. The self-modulated proton bunch resonantly excites a large amplitude plasma wave (wake field), which can be used for acceleration of plasma electrons. Analytical expressions for the linear growth rate and the number of exponentiations are given. We use the full three-dimensional particle-in-cell (PIC) simulations to study the beam self-modulation and the transition to the nonlinear stage. It is shown that the self-modulation of the proton bunch competes with the hosing instability which tends to destroy the plasma wave. A method is proposed and studied through PIC simulations to circumvent this problem which relies on the seeding of the self-modulation instability in the bunch.

Plasma based particle acceleration has made a significant progress over the last years. Since the first observations of the laser-driven wake field acceleration in the bubble regime [1], the electron bunch energies now routinely reach the near-GeV energies [2]. At the same time, the electron beam-driven plasma based acceleration experiments done at the Stanford Linear Accelerator (SLAC) have demonstrated energy gains of 42 GeV, and thus doubling the energy of the SLAC beam in a meter-scale plasma wake field accelerator [3].

However, on the path towards teraelectronvolt (TeV) energy range, one still has to solve many problems. The laser technology has to be improved a lot in terms of power, stability and repetition rate. On the other hand, the electron beam-driven wake field acceleration has the fundamental problem of the transformer ratio limit [4]. It arises because electric field of the same order drags the driver as the one that accelerates the witness beam, the maximum achievable energy gain in one single stage is limited by the energy of particles in the driver beam.

Recently, it has been proposed to harness a teraelectronvolt (TeV) proton bunch driver such as the one available from the LHC (Large Hadron Collider), at CERN (European Organization for Nuclear Research) for the TeV regime of electron acceleration in plasmas [5, 6]. Notwithstanding that the scheme demands for a short - of a sub-millimeter length - proton driver bunch, all the presently available proton bunches are much longer, usually tens of centimeter long. The bunches, however, have an excellent longitudinal emittance that would, in principle, allow to rotate them in the phase space [5, 7]. Yet, implementation of such phase rotation remains a challenge.

The available long proton bunches at the CERN cannot generate a high amplitude wake field directly. Indeed, for an efficient excitation of wakefields, the proton bunch length must be close to the plasma wavelength $\lambda_p = 2\pi c/\omega_p$, where $c$ is the speed of light and $\omega_p^2 = 4\pi n_e e^2/m_e$ is the square of the plasma frequency; $n_e$ is the plasma electron density, $e$ and $m_e$ are the electron charge and mass respectively. In a simplest way, plasma electrons can be understood as an ensemble of oscillators oscillating at plasma frequency. To enforce the resonant swinging of these oscillators, the driver must contain a Fourier component close to the plasma frequency. A short dense driver has a broad spectrum and thus generates the wake field efficiently. A long driver with the Gaussian temporal profile $\sim \exp\left(-t^2/\sigma_t^2\right)$ and $\sigma_t \omega_p \gg 1$ contains an exponentially small Fourier component at the plasma frequency and cannot be used directly to excite a plasma wave. The situation will change, however, if the long driver is modulated at the plasma frequency. Then, even a tenuous proton beam can excite a high amplitude plasma wave.

In this Letter, we develop an analytical model for the self-modulated regime of the proton bunch acceleration of electrons along the lines of the self-modulated laser wake-field accelerator (SMLWFA) concept [8]. The underlying physical mechanism is very much the same as in the SML-WFA concept. A long proton bunch $(L > \lambda_p)$ generates a wake within its body, which modulates the bunch itself, leading to the positive feedback and unstable modulation of the whole bunch along the bunch propagation direction. This self-modulation splits the long proton beam into ultra-short bunches of length $\sim \lambda_p$, which resonantly drive the plasma wake. This plasma wake can be used to accelerate electrons, just as in the SMLWFA concept.

The self-modulation of the proton bunch essentially occurs due to the action of the transverse wakefields on the bunch itself. This is different than the modulation caused by the electrostatic two-stream instability, which arises due to relative streaming between the proton bunch and the background plasmas, and causes excitation of the longitudinal fields. Due to the very anisotropic response of

4relativistic beam particles to longitudinal and transverse forces, modulation caused by the longitudinal forces becomes very inefficient in the case of a relativistic driver. This is precisely the reason that the transverse instabilities, such as the self-modulation instability, have higher growth rates than the longitudinal two-stream instability. In the context of a proton beam modulation, the two-stream instability is analogous to the forward Raman scattering of the laser pulse in laser driven wakefield acceleration. Although, the field of the beam-plasma instabilities is very rich, the analysis of the self-modulation instability of long proton bunches is conspicuously missing from the literature till now.

We begin with the analytical theory of the beam self-modulation based on the beam-envelope approach. Asymptotic expressions for the growth rate are obtained, and a simple semi-analytical code is developed to simulate the early stage of the instability. To substantiate the analytical results, we perform also fully electromagnetic three-dimensional particle-in-cell (3D-PIC) simulations. Finally, we dwell upon the competition of self-modulation instability with other instability such as the hose instability [9, 10] of the proton bunch, and evaluate a proposal to alleviate the effect of hosing instability, by PIC simulations, to produce an efficient excitation of wakefield by proton beams.

Following the approach of Ref. [11], we can write down the two-dimensional expressions for the wakefields of an axi-symmtric beam driver of an arbitrary profile, by utilizing the Euler variables $\xi = \beta_0 ct - z$, $\tau = t$, where $\beta_0 = v_z/c$ ($v_z$ is the velocity of the bunch), and assuming the quasi-static approximation ($\partial_\tau \simeq 0$) for the beam driver. Inside the body of a long proton bunch ($0 < \xi < L$), these read as

$$E_z(r,\xi) = 4\pi k_p^2 \int_0^\xi \int_0^\infty r' dr' \rho(r',\xi') I_0(k_p r_<) K_0(k_p r_>)$$
$$d\xi' f(\xi') \cos k_p(\xi - \xi'), \quad (1)$$

$$W_\perp(r,\xi) \simeq (E_r - B_\theta)(r,\xi) = 4\pi k_p \int_0^\xi \int_0^\infty r' dr' \, \partial_{r'} \rho(r',\xi')$$
$$I_1(k_p r_<) K_1(k_p r_>) \, d\xi' f(\xi') \sin k_p(\xi - \xi'), \quad (2)$$

where $\rho(r,\xi) = \rho_0 \psi(r) f(\xi)$ is the charge density of the bunch, $I_{0(1)}$ and $K_{0(1)}$ are the modified Bessel functions of order 0(1), $r_< = \min(r, r')$ and $r_> = \max(r, r')$, $k_p = \omega_p/c$ is the background plasma wave number, $L$ is the length of the bunch in the $\hat{z}$−direction, and we have assumed $\beta_0 \approx 1$.

The transverse wakefield acting on a long proton bunch, with a Heaviside step function profile ($\psi(r) = \Theta(r_b - r)$, $r_b$ being the radius of the beam-envelope) in radial direction and an arbitrary profile $f(\xi)$ in $\xi$ is written as

$$W_\perp(r,\xi) = -4\pi\rho_0 \int_0^\xi r_b(\xi') I_1\{k_p r(\xi')\} K_1\{k_p r_b(\xi')\} f(\xi')$$
$$k_p \sin k_p(\xi - \xi') d\xi', \quad (3)$$

where $\rho_0 = n_b e$ is the charge density of the proton bunch. Both the beam-envelope radius $r_b = r_b(\xi)$ and radial coordinate $r = r(\xi)$ are functions of $\xi$ on account of pinching caused by the wakefield on the beam.

The equation for the beam-envelope becomes then

$$\frac{\partial^2 r_b}{\partial \tau^2} - \frac{\mathcal{M}^2}{r_b^3} = -\frac{\omega_b^2}{\gamma_0} \int_0^\xi r_b(\xi') I_1\{k_p r_b(\xi')\} K_1\{k_p r_b(\xi')\}$$
$$f(\xi') k_p \sin k_p(\xi - \xi') d\xi', \quad (4)$$

where $\gamma_0 = (1 - \beta_0^2)^{-1/2}$ is the relativistic Lorentz factor of the beam, $\omega_b^2 = 4\pi\rho_0 e/m_b$ is the square of the non-relativistic beam plasma frequency of the proton bunch, $m_b$ being the mass of the beam particle. The constant $\mathcal{M}$ arises from the integration of the $\theta$-component of the equation of motion for the beam electrons yielding the angular momentum constant, and is associated with the transverse emittance of the beam [12]. For the demonstration of the self-modulation instability of a proton beam, we consider a thin beam ($k_p r_b \ll 1$) with a Heaviside step function profile ($f(\xi') = \Theta(\xi')$), and take $\mathcal{M} = \omega_{\beta 0} r_{b0}^2$, where $\omega_{\beta 0}^2 = \omega_b^2/2\gamma_0$ and $r_{b0}$ is the initial radius of the beam. The beam-envelope equation, in normalized coordinates ($r_b = r_b/r_{b0}$, $\tau = \omega_{\beta 0}\tau$, $\xi = k_p \xi$), reads as

$$\frac{\partial^2 r_b(\xi)}{\partial \tau^2} - \frac{1}{r_b^3(\xi)} = -\int_0^\xi r_b(\xi') \sin(\xi - \xi') d\xi'. \quad (5)$$

On perturbing Eq.(5) about the equilibrium radius $r_b = 1 + \delta r_b$, and assuming $\delta r_b = \delta \hat{r}_b \exp(i\xi)$, $|\partial \delta \hat{r}_b/\partial \xi| \ll |\delta \hat{r}_b|$, we obtain $(\partial_\xi^2 + 1)(\partial_\tau^2 + \Delta)\delta \hat{r}_b = -\delta \hat{r}_b$, where $\Delta = 3$. We assume the perturbation of the form $\delta \hat{r}_b(\xi,\tau) \sim \exp(i\delta\omega\tau - ik\xi)$, and obtain the dispersion relation $D \equiv (k^2 - 1)(\delta\omega^2 - \Delta) = -1$. The dispersion relation gives two complex $k$ roots (one in upper-half, and another in lower half of the complex-$k$ plane) for real $\delta\omega$ ($\sqrt{\Delta} < \delta\omega < \sqrt{1+\Delta}$). When $\Im(\delta\omega) \to \infty$, (i.e. $|\delta\omega| \to \infty$), the roots of the dispersion relation reach the real $k$-axis, thus confirming the presence of convective instability [13]. For complex $k$, the perturbation acquires the asymptotic form $\delta \hat{r}_b(\xi,\tau) \propto \exp(-ik_r\xi)\exp(k_i\xi)$, where $k_r$ and $k_i$ are the real and imaginary parts of the complex-$k$ root. For $k_i > 0$, the perturbation grows spatially in $\xi > 0$ direction, while for $k_i < 0$ (lower half $k$-plane) it grows spatially in $\xi < 0$ direction, thus representing a spatially amplifying wave.

One can derive asymptotic relations for the instability by following the approach of Bers [14]. For sufficiently late times, $\tau > L_e$, where $L_e \sim 1/\Gamma$ is the $e$-folding length, and $\Gamma$ is the growth rate of the instability, we solve this dispersion relation by letting $\delta\omega' = \delta\omega - \upsilon k$, where $\upsilon = \xi/\tau$, and setting $D(\delta\omega', k) = 0$ and $\partial D(\delta\omega', k)/\partial k = 0$ while keeping $\delta\omega'$ constant. This gives $k(\delta\omega^2 - \Delta)^2 = \delta\omega\upsilon$. We write $\delta\omega = \delta\omega_1 \pm \sqrt{\Delta}$. For $|\delta\omega_1| \gg \sqrt{\Delta}$ (early-time asymptote), we get $\Gamma_e = (\xi/2\tau)^{1/2}$. The number of exponentiation is given by $N_{ee} = \Gamma_e \tau = (\xi\tau/2)^{1/2}$. For $|\delta\omega_1| \ll \sqrt{\Delta}$ (late-time asymptote), we get the growth rate as $\Gamma_l = 3^{1/3}(\xi/\tau)^{2/3}/4$. The number of exponentiation is given by $N_{el} = \Gamma_l \tau = 3^{1/3}(\xi^2\tau)^{1/3}/4$. The expressions for number of exponentiations in dimensional units can be written as

$$N_{ee} = \left(\frac{\alpha}{2\sqrt{2\gamma_0}} k_p \xi \omega_p \tau\right)^{1/2}, \quad (6)$$

$$N_{el} = \frac{3^{1/3}}{4}\left(\frac{\alpha}{\sqrt{2\gamma_0}} k_p^2 \xi^2 \omega_p \tau\right)^{1/3}, \quad (7)$$

where $\alpha = \sqrt{(n_b/n_e)(m_e/m_b)}$.

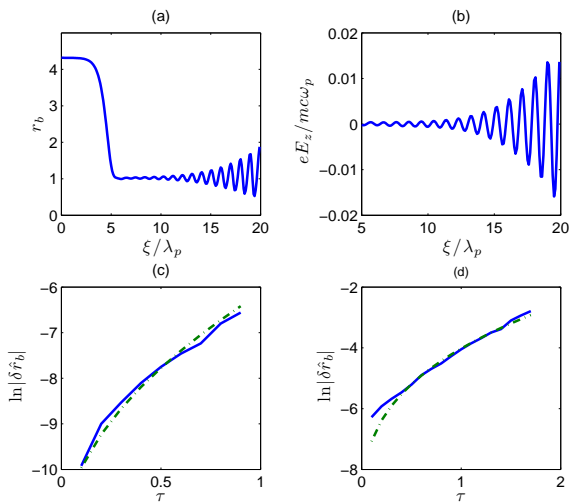

FIG. 1: The beam-envelope radius evolution with $\xi/\lambda_p$ at a time $\tau = 4.3$ showing the self-modulation instability of the beam (subplot(a)). The on-axis axial field generated by the beam (subplot(b)). Comparison of the amplitude of the beam's radius perturbations from Eq.(5) (solid lines) with the asymptotic relations (dash-dot lines).

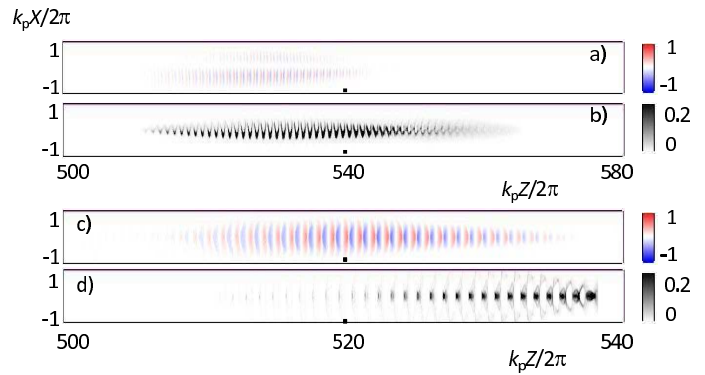

FIG. 2: 3D PIC simulation results for the smooth Gaussian beam (a), (b), and the hard cut half-Gaussian (c), and (d). The frames (a) and (c) show the accelerating wake field $E_z$, while the frames (b) and (d) show the beam density distribution.

We have solved Eq.(5) numerically to demonstrate the self-modulation instability of the beam-envelope for a gently rising beam density profile. Subplot (a) of Fig.1 shows the self-modulation instability of the beam-envelope with $\xi/\lambda_p$. The beam's head, $\xi/\lambda_p = 0$, is diffracting. Beam density has smoothly rising profile which flattens at $\xi/\lambda_p = 10$. The initial radius of the beam is $r_{b0}/\lambda_p = 0.1$. The boundary conditions are $r_b(\xi,0) = 1, \partial r_b(\xi,0)/\partial\tau = 0$. The self-modulation instability leads to the generation of strong axial field, depicted in subplot (b). The self-modulation of the beam grows in accordance with both the early-time (subplot (c)) and the late-time (subplot(d)) asymptotes, describing excellent agreements with the analytical scalings.

The propagation of the proton beam in plasmas also suffers from the onset of the hose instability which, in collisionless limit, is also known as the transverse two-stream instability [9, 10]. For a non-axisymmetric beam ($\partial_\theta \neq 0$), the hose instability is a concern. For a perfectly axi-symmetric beam, such as one assumed in our calculations, it doesn't occur. The temporal growth rate of the hose instability of a focused electron beam could be comparable to or less than the growth rate of the self-modulation instability. In a uniform plasma and in the limit of thick plasma skin depth $k_p r_b \ll 1$, the long-time asymptotic of the instability scales as $\Gamma_h \propto (k_\beta z'/\omega_p \tau')^{2/3}\omega_p$, where $k_\beta = k_b/\gamma_0^{1/2}$, $k_b = \omega_b/\upsilon_z$ being the beam betatron wave number, $\tau' = t - z/\upsilon_z$, $z' = z$ [9]. The ratio of growth rates for two instabilities scales as $\Gamma_l/\Gamma_h \propto (1/\alpha)^{1/3}(\gamma_0/2)^{1/6} \gg 1$, while $\alpha \ll 1$. Yet, it has time to develop and if present, can severely affect the wakefield. Thus, there is a big concern that the hosing instability occurring simultaneously with the beam self-modulation may destroy the plasma wake field. One of the possibilities to circumvent this problem is to pre-seed the self-modulation instability, so that the beam self-modulation does not have to grow from noise. This seeding would greatly increase both the shot-to-shot reproducibility and the quality of the wake field. Also, the development of the high amplitude wake field will require much shorter propagation distance, which will further limit the growth of the hose-instability. The seed of the self-modulation instability can be accomplished either by a short driver in front of the proton beam, or by a modification of the proton beam itself [15]. The latter can be achieved by cutting away the leading part of the proton beam in a "dogleg" device [16].

Because the hosing instability is not a part of the beam envelope analytic description (4), we have to rely here on 3D PIC simulations, which were performed using the code VLPL [17]. We simulate two cases. In the both cases, the proton beam is assumed to have 24 GeV energy. In the first case, the beam has the smooth 3D Gaussian density profile $n_{b1}(r,z) = n_b^0 \exp\left(-z^2/\sigma_z^2\right) \exp\left(-r^2/\sigma_r^2\right)$, where $r^2 = x^2 + y^2$. In the second case the beam was hard cut in the middle, $n_{b2}(r,z) = n_b^0 \Theta(-z) \exp\left(-z^2/\sigma_z^2\right) \exp\left(-r^2/\sigma_r^2\right)$, where $\Theta(z)$ is again the Heaviside step-function. The plasma is assumed to be singly ionized lithium. The electron density $n_e$ of the plasma is 25 times higher than the peak beam density, i.e. $n_b^0 = 0.04 n_e$. The beam in the simulation has an r.m.s transverse momentum spread of $\sqrt{<p_\perp^2>}/m_i c = 4.5 \times 10^{-3}$. The beam radius measured in the plasma skin lengths was $k_p \sigma_r = 2\pi \cdot 0.25$. The beam length was $k_p = 2\pi \cdot 20$. The beam has a relativistic Lorentz factor $\gamma_0 = 25$.

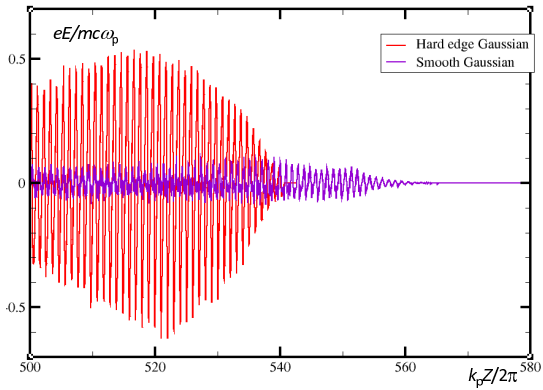

FIG. 3: Direct comparison of the on-axis wake fields generated after the propagation length of 500 plasma periods for the both cases, the smooth Gaussian and the hard cut half-Gaussian.

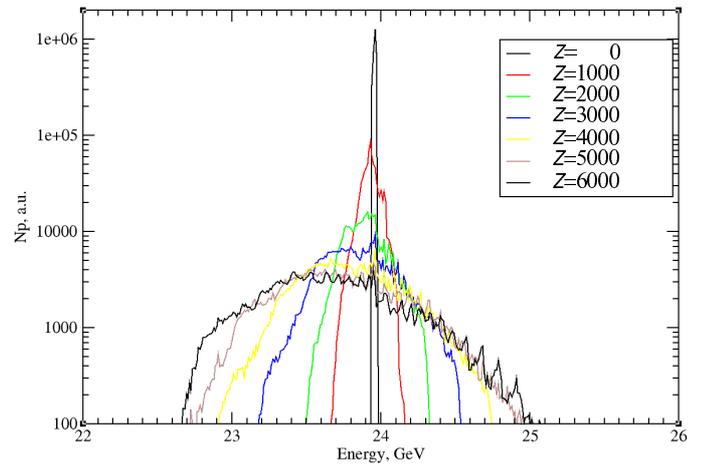

FIG. 4: Energy spectrum evolution of the hard cut half-Gaussian proton beam. Spectrum broadening and acceleration of some protons up to 1 GeV is observed.

The simulation results showing the on-axis generated wake field and the self-modulated beam are given in Fig. 2 in a $ZX-$ plane. The frames (a) and (b) show the smooth Gaussian beam case, while the frames (c) and (d) correspond to the second case of the hard cut Gaussian beam. One sees that in the first case when the both instabilities have to grow from noise, the hosing instability competes with the self-modulations. Although the beam is split into small beamlets, these density perturbations are located slightly off-axis, and even some filamentation can be observed. On the contrary, when the self-modulation instability is seeded by the hard cutting of the beam in the second case, the wake field is very regular and axisymmetric. To compare the amplitude of the excited wake fields in the both cases, we plot the on-axis accelerating field in Fig. 3. The wake field for the hard cut Gaussian beam is approximately 5 times larger than that for the smooth Gaussian case. The maximum accelerating field normalized to the wave breaking field

[16] is $eE_z/m_e c \omega_p \approx 0.6$. Thus, it is close to the nonlinear regime. Fig. 4 shows the energy spectrum evolution of the proton beam. One observes the spectrum broadening as the wake field is developing. Finally, some protons acquire energy gain of nearly 1 GeV.

In summary, we have demonstrated the self-modulation instability of a long proton bunch, which can resonantly excite the plasma wave needed for the teraelectronvolt regime of electron acceleration. We have analyzed the seeding of the self-modulation instability in order to alleviate the effects of the hose instability on the plasma wave. The seeding was accomplished by hard cutting a Gaussian beam. The simulation results of the hard cut Gaussian beam show that in this case the generated wakefield acquires high values and provides a substantial acceleration of the plasma electrons. Further, the combination of seeding of the self-modulation instability and by suitably restricting the beam propagation length in plasmas, one can hope to excite large amplitude plasma waves, and this could pave way for the successful realization of the teraelectronvolt regime of electron acceleration scheme.

We acknowledge useful discussions with Prof. Ch. Joshi, Prof. R. Bingham, Prof. P. Muggli, Prof. A. Caldwell, Dr. Guoxing Xia and Dr. W. Lu. We also thank Daniel an der Brügge for help with the semi-analytical code.

---

* Electronic address: kumar@tp1.uni-duesseldorf.de
[1] A. Pukhov and J. Meyer-ter Vehn, Applied Physics B: Lasers and Optics **74**, 355 (2002), URL http://dx.doi.org/10.1007/s003400200795; J. Faure, Y. Glinec, A. Pukhov, S. Kiselev, S. Gor-

dienko, E. Lefebvre, J. P. Rousseau, F. Burgy, and V. Malka, Nature **431**, 541 (2004), URL http://dx.doi.org/10.1038/nature02963; C. G. R. Geddes, C. Toth, J. van Tilborg, E. Esarey, C. B. Schroeder, D. Bruhwiler, C. Nieter, J. Cary, and W. P. Leemans, Nature **431**, 538 (2004), URL http://dx.doi.org/10.1038/nature02900; S. P. D. Mangles, C. D. Murphy, Z. Najmudin, A. G. R. Thomas, J. L. Collier, A. E. Dangor, E. J. Divall, P. S. Foster, J. G. Gallacher, C. J. Hooker, et al., Nature **431**, 535 (2004), URL http://dx.doi.org/10.1038/nature02939.
[2] W. P. Leemans, B. Nagler, A. J. Gonsalves, C. Toth, K. Nakamura, C. G. R. Geddes, E. Esarey, C. B. Schroeder, and S. M. Hooker, Nature Phys. **2**, 696 (2006), URL http://dx.doi.org/10.1038/nphys418; S. Kneip, S. R. Nagel, S. F. Martins, S. P. D. Mangles, C. Bellei, O. Chekhlov, R. J. Clarke, N. Delerue, E. J. Divall, G. Doucas, et al., Phys. Rev. Lett. **103**, 035002 (2009).
[3] I. Blumenfeld, C. E. Clayton, F.-J. Decker, M. J. Hogan, C. Huang, R. Ischebeck, R. Iverson, C. Joshi, T. Katsouleas, N. Kirby, et al., Nature **445**, 741 (2007), URL http://dx.doi.org/10.1038/nature05538.
[4] E. Esarey, P. Sprangle, J. Krall, and A. Ting, IEEE Trans. Plasma Sci. **24,**, 252 (1996), URL http://dx.doi.org/10.1109/27.509991.
[5] A. Caldwell, K. Lotov, A. Pukhov, and F. Simon, Nat Phys **5**, 363 (2009), URL http://dx.doi.org/10.1038/nphys1248.
[6] K. Lotov, Phys. Rev. ST - Accel. Beams ((to be published)).
[7] L. Evans and P. B. (editors), Journal of Instrumentation **3**, S08001 (2008), URL http://stacks.iop.org/1748-0221/3/S08001.
[8] E. Esarey, J. Krall, and P. Sprangle, Phys. Rev. Lett. **72**, 2887 (1994).
[9] D. H. Whittum, Phys. Fluids B **5**, 4432 (1993), URL http://link.aip.org/link/?PFB/5/4432/1.
[10] J. Krall and G. Joyce, Physics of Plasmas **2**, 1326 (1995), URL http://link.aip.org/link/?PHP/2/1326/1; A. A. Geraci and D. H. Whittum, Phys. Plasmas **7**, 3431 (2000).
[11] R. Keinigs and M. E. Jones, Phys. Fluids **30**, 252 (1987).
[12] E. P. Lee and R. K. Cooper, Part. Accel. **7**, 83 (1976).
[13] E. M. Lifshitz and L. P. Pitaevskii, *Physical Kinetics*, vol. 10 of *Course of Theoretical Physics* (Butterworth-Heinemann Ltd., 1981), 1st ed.
[14] A. Bers, *Basic Plasma Physics I* (North-Holland Publishing Company, 1983), vol. 1 of *Handbook of Plasma Physics*, chap. 3.2.
[15] K. Lotov, in *6th European Particle Accelerator Conference* (Stockholm, 1998), pp. 806–808.
[16] P. Muggli, V. Yakimenko, M. Babzien, E. Kallos, and K. P. Kusche, Phys. Rev. Lett. **101**, 054801 (2008).
[17] A. Pukhov, J. Plasma Phys. **61**, 425 (1999).